\documentclass[aps,prl,twocolumn,superscriptaddress,showpacs]{revtex4}

\usepackage{graphics}

\bibliography{20}
\begin{document}
\title{Conservation of Angular Momentum, Transverse Shift, and Spin Hall Effect
in Reflection and Refraction of Electromagnetic Wave Packet}

\author{Konstantin Yu. Bliokh}

\affiliation{Institute of Radio Astronomy, 4 Krasnoznamyonnaya st., Kharkov, 61002, Ukraine}

\affiliation{Department of Physics, Bar-Ilan University, Ramat Gan, 52900, Israel}

\author{Yury P. Bliokh}

\affiliation{Physics Department, Technion-Israel Institute of Technology, Haifa, 32000, Israel}

\begin{abstract}
We present a solution to the problem of reflection/refraction of
a polarized Gaussian beam on the interface between two
transparent media. The transverse shifts of the beams' centers of
gravity are calculated. They always satisfy the total angular
momentum conservation law for beams, however, in general, do not
satisfy the conservation laws for individual photons in
consequence of the lack of the ``which path'' information in a
two-channel wave scattering. The field structure for the
reflected/refracted beam is analyzed. In the scattering of a
linearly-polarized beam, photons of opposite helicities are
accumulated at the opposite edges of the beam: this is the spin
Hall effect for photons, which can be registered in the
cross-polarized component of the scattered beam.
\end{abstract}
\pacs{42.90.+m, 42.25.Gy, 42.15.-i, 42.25.Ja}

\maketitle

\paragraph{Introduction. ---}Reflection/refraction of plane electromagnetic waves
at the interface between two homogeneous transparent media is
described by the Snell law and the Fresnel formulas \cite{1}.
However, in the case of the localized wave packets (or beams) the
Snell law as well as the Fresnel formulas (as shown below) give no
exact description of their refraction and reflection. First, the
reflected packet undergoes a short longitudinal shift in the
reflection plane: this is the Goos--H\"{a}nchen effect \cite{2},
which is not a subject of this work. Besides, a circularly- (or
elliptically-) polarized incident packet experiences a {\em
transverse shift} (TS) and leaves the plane of incidence when
refracting or reflecting. This effect was originally predicted by
Fedorov \cite{3} and since that time has been discussed in a
number of papers, both theoretical \cite{4, 5, 6, 7} and
experimental \cite{8, 9, 10}.

TS plays a fundamental role in electrodynamics: this phenomenon is
responsible for the conservation of the {\em total angular
momentum} (TAM) of an electromagnetic beam, including the
intrinsic (spin) part \cite{5, 7}. For a smoothly inhomogeneous
medium this effect represents the optical Magnus effect \cite{6,
9, 11} or the recently discovered topological spin transport (spin
Hall effect) of photons \cite{7, 12}. However, in spite of a long
period of research, up until now the ultimate answer as to the
magnitude of TS, along with the correct wording of the TAM
conservation law for an electromagnetic beam, is not found: almost
all papers \cite{3, 4, 5, 6, 7} result in different answers.

In this Letter we propose an exact solution to the problem of
reflection/refraction of an arbitrary polarized Gaussian beam in the framework of classical electrodynamics. This enables us to evaluate TSs of the
centers of gravity of the scattered beams, to determine the TAM
conservation law that governs the process, and to analyze the
field structure in the beams, which reveals the spin Hall effect
for photons. It is shown that the mixing of classical and quantum arguments can lead to an incorrect determination of the beam TS.

\paragraph{Angular momentum conservation laws. ---}TAM of a polarized
electromagnetic wave packet, $\mathbf{J}$, consists of the orbital
momentum $\mathbf{L}$ and the intrinsic (spin) momentum
$\mathbf{S}$. The TAM density (TAM of one photon) can be
represented as $\mathbf{j} = \mathbf{r} \times \mathbf{k} +
(e|\sigma _3 |e)\mathbf{k}/k$ ($\hbar  = c = 1$) \cite{7}. Here
$\mathbf{r}$, $\mathbf{k}$, and $|e)$ are the radius-vector, wave
vector, and two-component polarization vector of the wave packet
center ($|e)$ is written in the basis of circular polarizations,
i.e. helicity basis); $\sigma _3  = {\rm diag} \left( {1, - 1}
\right)$ is the Pauli matrix. TAM is related to the TAM density as
$\mathbf{J} = N\mathbf{j}$, where $N = W/\omega$ is the number of
photons in the packet, $W$ is the total field energy of the beam,
and $\omega$ stands for the frequency (we consider a monochromatic
packet). When a wave packet is scattered on the interface $z=0$
between two homogeneous media, the normal to the surface component
of TAM is conserved owing to the axial symmetry of the problem:
$J_z^{(i)} = J_z^{(r)} + J_z^{(t)}$. From here on, the
superscripts $(i)$, $(r)$, and $(t)$ correspond to the incident,
reflected, and refracted wave packets, respectively. The energy
conservation law results in the conservation of the total number
of photons: $N^{(i)} = N^{(r)} + N^{(t)}$. Taking into account
that $W \propto \varepsilon \left| \mathbf{E} \right|^2 V$
($\varepsilon$ stands for the permittivity, $\mathbf{E}$ is the
electric field in the wave packet, and $V$ is the packet volume)
and that the volume of the packet varies as $V \propto n^{-1} |\cos
\vartheta|$ in the course of refraction ($n = \sqrt {\varepsilon
\mu}$ is the refraction index, while $\vartheta$ is the angle
between $\mathbf{k}$ and $z$-axis), the conservation law for the
$z$-component of TAM reads \cite{5}:

\begin{equation}\label{eq1}
j_z^{(i)}  =  R^2 j_z^{(r)}  + T^2 \frac{{n_2 \mu _1 \cos
\theta'}} {{n_1 \mu _2 \cos \theta}}j_z^{(t)}.
\end{equation}
Here $R,T = \left| {\mathbf{E}^{(r,t)} } \right|/\left|
{\mathbf{E}^{(i)} } \right|$  are the Fresnel
reflection/refraction coefficients for plane waves, subscripts 1
and 2 refer to parameters of the first and the second medium,
$\mu$ stands for the permeability, and we denote
$\vartheta^{(i)}=\theta, \vartheta^{(t)}=\theta', \vartheta^{(r)}
= \pi - \theta$ (Fig. 1). Eq. (\ref{eq1}) constitutes the main TAM
conservation law for a wave packet; it has a classical nature and follows immediately from the Maxwell equations \cite{5}.

\begin{figure}[t]
\centering \scalebox{0.9}{\includegraphics{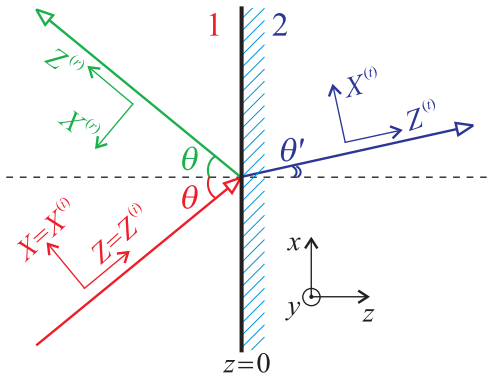}} \caption{(Color online) The
scheme of the wave reflection and refraction with beam coordinates
used in the text.}

\label{Fig1}
\end{figure}

Eq. (\ref{eq1}) is inadequate for determining the shifts of the
centers of gravity for the reflected and refracted wave packets
since one equation contains two unknown terms. Another approach
has been suggested in \cite{7}, where the authors consider the
wave packet scattering as a set of individual
reflection/refraction events of isolated photons. In this case
each photon finds itself either in the medium 1 (is reflected) or
in the medium 2 (is refracted); hence two TAM conservation laws
take place for one photon:

\begin{equation}\label{eq2}
j_z^{(i)}  = j_z^{(r)},\hspace{5mm}j_z^{(i)}  = j_z^{(t)}.
\end{equation}
Eqs.(\ref{eq2}) determined TSs of the wave packets in \cite{7};
numerical simulation for a circularly-polarized incident wave
packet has shown a good agreement with the theory.

At the same time, Eqs. (\ref{eq2}) do {\em not} always satisfy the
main conservation law (\ref{eq1}). Eqs. (1) and (2) coincide only
in particular cases: e.g. in the case of total internal
reflection, where $R=1$, $T=0$, or in the case of small contrast
between two media, $\left| {n_2  - n_1 } \right| \equiv \delta n
\ll 1$. In the latter case $T^2 \frac{{n_2 \mu _1 \cos \theta '}}
{{n_1 \mu _2 \cos \theta }} = 1 + O\left( {\delta n^2 } \right)$,
$R^2 = O\left( {\delta n^2 } \right)$ and Eq. (\ref{eq1}) is
equivalent to the second Eq. (\ref{eq2}) in the linear
approximation in $\delta n$, i.e. in the geometrical optics
approximation \cite{6, 7, 11, 12}. This clarifies the fact that
$j_z$  is an exact integral of motion for the modified equations
of geometrical optics in an axially symmetrical medium \cite{7}.
Below is shown that Eqs. (\ref{eq2}) are consistent with Eq.
(\ref{eq1}) for a circularly polarized initial beam. However, this
is not true in the general case of an elliptically polarized beam.
We will demonstrate that the scattering of the Gaussian
electromagnetic beam in all cases satisfies Eq. (\ref{eq1}) and not
Eqs. (\ref{eq2}). The fallacy of the Eqs. (\ref{eq2}) stems
apparently from the quantum-mechanical approach \cite{7}, which is
based on the events for individual photons. The point is that
classical electrodynamics describes a multiphoton interference
pattern. By invoking the conservation laws (\ref{eq2}) we invoke
thereby a ``which path'' information, which destroys, as is well
known from quantum mechanics, the interference pattern in the
multiphoton scattering. Thus Eqs. (\ref{eq2}) are
suitable for describing the scattering process of individual
photons; however in the generic case they are inapplicable in the
scattering of classical wave packets \cite{125}.

\paragraph{Gaussian beam reflection and refraction. Transverse shift. ---}The electric field of the
wave packet incident in the plane $\left( {x,z} \right)$ can be
represented in the form of a polarized Gaussian beam:

\begin{equation}\label{eq3}
\mathbf{E}^{(i)}  = A\frac{{\mathbf{e}_X  + m\left( {\mathbf{e}_y
- yB\mathbf{e}_Z } \right)}} {{\sqrt {1 + \left| m \right|^2 }
}}\exp \left( {ikZ + \frac{{ikBy^2 }} {2}} \right).
\end{equation}
Here we use a reference frame $XyZ$  associated with the beam
(Fig. 1), $\mathbf{e}_{X,y,Z}$ are its unit vectors, the complex
value $m$ is related to the beam polarization (the polarization in
the beam center, $y=0$, is characterized by the polarization
vector $\left( {\mathbf{e}_X + m\mathbf{e}_y } \right)/\sqrt {1 +
\left| m \right|^2 } $, or in the helicity basis, $|e) = \left(
{\begin{array}{*{20}c}
   {1 - im}  \\
   {1 + im}  \\
 \end{array} } \right)/\sqrt {2\left( {1 + \left| m \right|^2 } \right)}
$), and the complex parameters $A = A\left( Z \right)$ and $B =
B\left( Z \right)$ vary along the beam in consequence of its
diffraction (standard solutions in a homogeneous media are
obtainable in the framework of the complex geometrical optics
\cite{13}). $A$ is the beam amplitude, while the real and
imaginary parts of $B$ are responsible for the phase front
curvature and the beam width, respectively.

For the sake of simplicity we assume that the beam (\ref{eq3}) is
confined in $y$ only, which enables us to consider only TS along
this coordinate and not the Goos--H\"{a}nchen effect. The
deviation from the center, $y \ne 0$, results in a small
longitudinal (in the $\mathbf{e}_Z$-direction) field component
proportional to $y$. Owing to this component, field (\ref{eq3})
satisfies the Maxwell equations for a homogeneous medium: ${\rm
div} \mathbf{E} = 0$, i.e. $\mathbf{E}$ is orthogonal to the local
wave vector $\mathbf{k}_{loc}$. It is the longitudinal field
component that is responsible for TS of the beam center in the
process of beam reflection and refraction. The representation of
the Gaussian polarized beam in the form of (\ref{eq3}) holds good
for sufficiently large distances $y$ until the wavelength is small
compared to the characteristic beam width and the radius of
curvature of its phase front: $\left| B \right|y \ll 1$.

The field of the reflected/refracted beam can be obtained from Eq.
(\ref{eq3}) supplemented by standard boundary conditions \cite{1}.
As a result of cumbersome but direct calculation, the fields
for all three beams, (i), (r), and (t), can be written in a
unified form:
\begin{eqnarray}
\label{eq4}
\mathbf{E}^a  = {A^a S^a \over\sqrt {1 + \left| m^a  \right|^2 }}\exp \left( ik^a Z^a  + {ik^a B^a y^2 \over 2}\right)\nonumber\\
\left\{ \left[ 1 + {m^a B^a y\over \rho ^a \sin \vartheta ^a }\left( \cos \theta  - \rho ^a \cos \vartheta ^a  \right) \right]\mathbf{e}_{X^a }\right.\hspace{3mm}\\
\left. + \left[ m^a  + {B^a y\over\sin\vartheta ^a }\left( \cos \vartheta ^a  - \rho ^a \cos
\theta  \right) \right]\mathbf{e}_y  - m^a B^a y\mathbf{e}_{Z^a }  \right\} \nonumber
\end{eqnarray}
where  $a=(i),(r),(t)$,  $S^{(i,r,t)}=1,R {\rm sgn} R_\parallel,T$, $\rho ^{(i,r,t)} =
1,\,R_ \bot  /R_\parallel ,\,T_ \bot  /T_\parallel $,  $m^a  =
\rho ^a m$ is the characteristic of a central polarization of the
corresponding beam, $k^{(i)}  = k^{(r)}  = k$,  $k^{(t)}  = kn_2
/n_1$, $A^{(i)} = A$,  $B^{(i)} = B$, $A^{(r,t)}$ and $B^{(r,t)}$
are determined from the boundary conditions $\left. {A^{(r,t)} }
\right|_{z = 0}  = \left. A \right|_{z = 0}$ and $\left.
{B^{(r,t)} } \right|_{z = 0}  = \left. B \right|_{z = 0}$. For all
beams the associated Cartesian coordinates  $X^a yZ^a$ and their
unit vectors $\mathbf{e}_{X^a,y,Z^a }$ are used (Fig. 1). In the
definitions above, we used the Fresnel coefficients for the plane waves whose electric vector is parallel/orthogonal to the incidence plane \cite{1}:
%\begin{eqnarray}\label{eq5}
$T_\parallel   = {2\varepsilon _1 n_2 \cos \theta }/ (\varepsilon
_2 n_1 \cos \theta  + \varepsilon _1 n_2 \cos \theta '),$
%\nonumber\\
$T_ \bot   = 2\mu _2 n_1 \cos \theta / (\mu _2 n_1 \cos \theta +
\mu _1 n_2 \cos \theta '),$
%\nonumber\\
$R_\parallel   = 1 - \cos \theta ' T_\parallel/\cos
\theta,\hspace{3mm}R_ \bot = T_ \bot   - 1\,.$
%\nonumber
%\end{eqnarray}

To determine the beams' centers of gravity, let us consider the
projection of the field (\ref{eq4}) onto its central polarization
vector \cite{6}:  $F = \mathbf{E}^a \left( {\mathbf{e}_{X^a }  +
m^a \mathbf{e}_y } \right)/\sqrt {1 + \left| {m^a } \right|^2 }$.
This value can be represented in the first approximation in
$\left| B \right|y \ll 1$ as $F = A^a S^a \exp \left[ {ik^a Z^a  +
{ik^a B^a \left( {y - \delta y^a } \right)^2 } / 2} \right]$,
where $\delta y^a$ is a complex value ($\delta y^{(i)} = 0$). Its
imaginary part is responsible for phenomena associated with a
phase front curvature (they are omitted in the present study
\cite{14}), while ${\rm Re}~\delta y^a = \Delta ^a $, represents
TS associated with the TAM conservation. Calculations yield

\begin{equation}\label{eq6}
\Delta ^a  = -{\cot \theta\over k}{{\rm Im}\, m\left( 1 + \rho
^{a2} - 2\rho ^a {\cos \vartheta ^a\over\cos \theta } \right)\over 1
+ \rho ^{a2} \left| m \right|^2 }~.
\end{equation}
Formulas for $T_{\parallel,\bot}$ and $R_{\parallel,\bot}$ have
been considered real, which excludes the case of total internal
reflection, where $\cos \theta ' = \sqrt {1 - \left( {n_1^2 /n_2^2
} \right)\sin ^2 \theta } $ becomes imaginary. Calculations for
the totally reflected beam read

\begin{equation}\label{eq7}
\Delta ^{(tot\,{\kern 1pt} r)} = -{2\cot \theta \over k}{{\rm Im}
m\left( 1 + {\rm Re} \rho ^{(r)} \right) + {\rm Re} m{\rm Im} \rho
^{(r)} \over 1 + \left| m \right|^2 }~.
\end{equation}

Let us verify now the agreement between Eqs. (\ref{eq6}),
(\ref{eq7}) and the TAM conservation laws (\ref{eq1}),
(\ref{eq2})). The $z$-component of the beam's TAM density equals
\cite{5,7} $j_z^a  = - \Delta ^a k^a \sin \vartheta^a + \frac{{2{\rm
Im} m^a }} {{1 + \left| {m^a } \right|^2 }}\cos \vartheta ^a $. By
substituting these values along with TSs (\ref{eq6}) and
(\ref{eq7}), and $R = \sqrt {\left| {R_\parallel } \right|^2  +
\left| {R_ \bot } \right|^2 \left| m \right|^2 } /\sqrt {1 +
\left| m \right|^2 }$, $T = \sqrt {\left| {T_\parallel } \right|^2
+ \left| {T_ \bot } \right|^2 \left| m \right|^2 } /\sqrt {1 +
\left| m \right|^2 }$ (or $T=0$ for the total internal reflection)
into (\ref{eq1}), we make sure that the TAM conservation law
(\ref{eq1}) is satisfied identically. At the same time, shifts
(\ref{eq6}) and (\ref{eq7}) satisfy the TAM conservation laws for
individual photons, Eqs. (\ref{eq2}), solely in the following
particular situations: A) the incident beam is linearly polarized,
${\rm Im}~m = 0$, and $\Delta^a = 0$; B) the incident beam is
circularly polarized, $m = \pm i$ (this explains a good agreement
of numerical simulation in \cite{7} with Eqs. (\ref{eq2})); C) the
case of total internal reflection, where just one scattering
channel exists and laws (\ref{eq1}) and (\ref{eq2}) are identical.
In all other cases shifts (\ref{eq6}) do not satisfy Eqs.
(\ref{eq2}).

\paragraph{Spin Hall effect of light. ---}Eq. (\ref{eq4}) enables one not only to
find TS of the centers of gravity but also to determine the field
structure in the reflected/refracted beam. One can see that the
reflected/refracted beam does not have the form of the Gaussian
beams (\ref{eq3}) shifted in accordance with Eqs. (\ref{eq6}) and
(\ref{eq7}). However, in the first approximation in  $\left| B
\right|y \ll 1$, field (\ref{eq4}) can be represented as a
superposition of two Gaussian circularly polarized beams like
(\ref{eq3}):  $\mathbf{E}^a  = \alpha ^ + \mathbf{E}^{a + }  +
\alpha ^ -  \mathbf{E}^{a - } $, where $\alpha ^ \pm   = \left( {1
\mp im^a } \right)/\sqrt {2\left( {1 + \left| {m^a } \right|^2 }
\right)} $ and
%\begin{eqnarray}\label{eq8}
$\mathbf{E}^{a \pm } = A^a S^a \frac{{\mathbf{e}_{X^a }  \pm
i\left( {\mathbf{e}_y  - yB^a \mathbf{e}_{Z^a } } \right)}}{{\sqrt
2 }}
%\nonumber\\
\exp \left[ {ik^a Z^a  + \frac{{ik^a B^a \left( {y - \delta^{a \pm }
} \right)^2 }} {2}} \right]$.
%\end{eqnarray}
Assume that the incident beam is linearly polarized with the
electric field parallel or orthogonal to the incidence plane:
$m_{\parallel , \bot } = 0,\infty$. In this case $\Delta ^a = 0$;
however, the shifts of the reflected and refracted partial beams
$\mathbf{E}^{a \pm }$ are nonzero and oppositely directed: $\delta
_\parallel ^{a \pm } =  \pm \left( {\cos \vartheta ^a  - \rho ^a \cos
\theta } \right)/k\sin \theta$ and $\delta _ \bot ^{a \pm } = \pm
\left( {\cos \vartheta ^a - {\rho ^a} ^{ - 1} \cos \theta }
\right)/k\sin \theta$ (here we do not consider the total internal
reflection case). This confirms the predicted earlier effect of
splitting of a beam of mixed polarization into two circularly
polarized beams in an inhomogeneous medium \cite{11, 12}. The
splitting is very small -- fractions of the wavelength;
nevertheless, it leads to new observable phenomena.

Indeed, the elliptical polarizations of opposite signs arise at
the opposite edges of the beam. (As a consequence, the beam as a
whole is depolarized, i.e. in contrast to the Fresnel formulas for
plane waves, the polarization state of the linearly polarized beam
changes after reflection/refraction and becomes mixed.) In the
approximation considered, the degree of the circular polarization is
proportional to $y$, i.e. it grows linearly with the distance from the
beam center. The initiation of elliptical polarizations at the
ends of a linearly polarized beam is a manifestation of the spin
Hall effect for photons: the photons of opposite helicities
accumulate at the opposite ends of the beam just as in the
recently discovered spin Hall effect for carriers in
semiconductors \cite{15,16}. It is interesting that this effect
for photons was predicted as early as 1965 in the paper of Costa
de Beauregard \cite{4}.

\begin{figure}[t]
\centering \scalebox{0.44}{\includegraphics{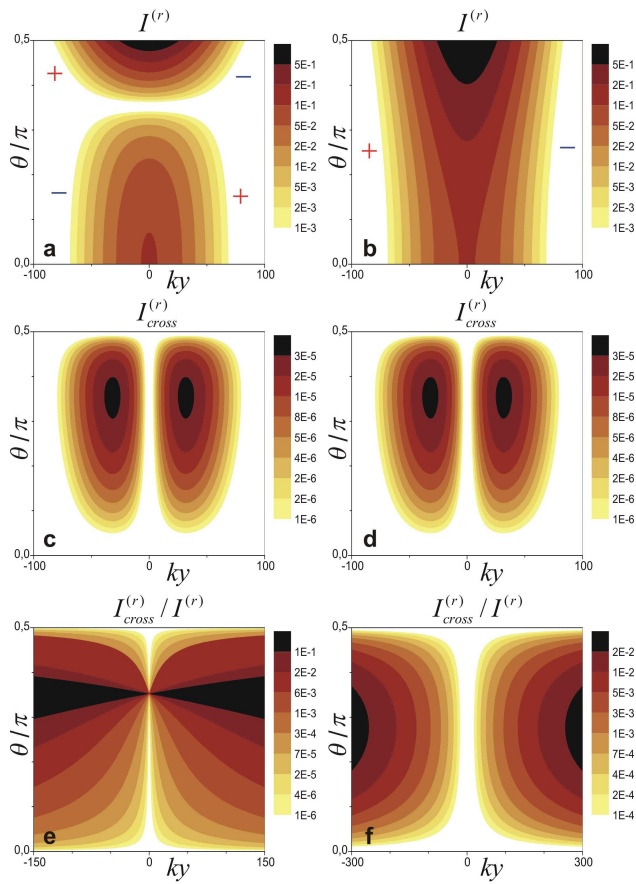}}
\caption{(Color online) Intensity of the reflected beam's field (a,b), and
absolute (c,d) as well as relative (e,f) intensity of
cross-polarized component in the beam via the transverse
coordinate and incidence angle at different polarizations of the
initial beam ($m=0$ for (a,c,e) and $m=\infty$ for (b,d,f)). The
signs of the elliptical polarizations at the different sides of
the beam are marked (a,b). Parameters are: $n_2 /n_1 = 2$, $\mu
_{1,2} = 1$, $A^{(r)} = 1$, and $k^{ - 1} B^{(r)} = i\,10^{ - 3}$
in the observation point, which corresponds to the beam's width
about 10 wavelengths.}

\label{Fig2 }
\end{figure}

The change in the polarization structure along with the splitting
of a linearly polarized beam can be observed experimentally by
measuring the cross-polarized field component of the
reflected/refracted beam (i.e. $E_{y}^{a}$ and $E_{X^a}^{a}$ for
$m=0$ and $\infty$, respectively). The intensity of the
cross-component, being closely related to the degree of the
circular polarization, is equal to $I_{cross}^{a} = I^a \left| {B^a
k^a \delta ^{a + } } \right|^2 y^2 \propto y^2 \exp \left( - {\rm
Im} B^a y^2\right)$, where $I^a = {\left| A^a \right|}^2 {S^a}^2
\exp \left( - {\rm Im} B^a y^2 \right)$ is the field intensity in
the beam. The relative cross-component intensity grows infinitely
with $y$: $I_{cross}^{a} /I^a \propto y^2 $. Fig. 2 presents the
distributions of absolute and relative cross-component intensities
in the reflected beam. The beam splitting is easily visible, while
the angle corresponding to the maximum of the absolute
cross-component intensity visually coincides with the angle
associated with the maximum TS of the circularly polarized beam
\cite{7}. A flip of the helicity (Fig. 2a) and a singularity in the relative cross-component intensity (Fig. 2e) for $m=0$  occur at the Brewster angle where in-plane component $E^{(r)}_{X^{(r)}}$ vanishes and changes its sign. With the characteristic parameters of
present-day optical polarizers and laser beams, one can look
forward to detect the cross-component like that in Fig. 2 and to register the spin Hall effect of photons.

\paragraph{Conclusion. ---}We have solved the problem of reflection/refraction of
an electromagnetic polarized Gaussian beam at the interface
between two homogeneous media. The transverse shifts of the
centers of gravity for the reflected and refracted beams have been
calculated. In all cases they satisfy the total angular momentum
conservation law for beams, but in the generic case do not satisfy the
conservation laws for individual photons because of the
fundamentally two-channel character of the process and the lack of
``which path'' information in classical electrodynamics. The field
structure for the reflected/refracted beam has been analyzed.
Initially linearly polarized beam splits into two circularly
polarized beams shifted in opposite directions. This causes the
rise of elliptical polarizations of opposite signs at the beam
edges, i.e. the spin Hall effect for photons. The effect can be
detected by measuring the split cross-component of the scattered
beam's field.
\begin{acknowledgments}
The work was supported by INTAS grant 03-55-1921.
\end{acknowledgments}

\paragraph{Note added. ---}Experimental works by Punko and Filippov with observation of the spin Hall effect of light described in the present paper came to our attention after publication of our work. N.N.~Punko and V.V.~Filippov, Pis'ma v ZhETF {\bf 39}, 18 (1984) [JETP Letters 39, 20 (1984)]; Opt. Spektrosk. {\bf 58}, 125 (1985).


\begin{references}
\bibitem{125}{When a single photon is scattered, it can be either reflected or refracted, the observer evidently has "which path" information, and Eq. (2) holds true. If a number of photons experience scattering simultaneously, the observer has no chance to establish a one-to-one correspondence between incident and reflected/refracted photons, the interference takes place and only the total angular momentum is conserved, Eq. (1).}
\bibitem{14}{A shift of the beam's center of gravity, which is related to ${\rm Im}~\delta y^a \propto {\rm Re}~m$,
can be excluded either by considering the beam in the point of
flat phase front or (not in the total internal reflection case) by
assuming ${\rm Re}~m = 0$. The latter means that the the
polarization ellipse of the incident beam is oriented along $X$
and $y$ coordinate axes.}
\end{references}
\end{document}